\documentclass[twocolumn,pra,psfig,showpacs,superscriptaddress]{revtex4-1}
\usepackage{amssymb}
\usepackage{amsfonts}
\usepackage{amsmath}
\usepackage{graphicx}

\begin{document}

\title{Approximative Wigner function for the Helium atom and dissipation}
\author{H. Dessano}\email{haradessano@gmail.com}
\affiliation{International Center for Condensed Matter Physics, Instituto de F\'{\i}sica, Universidade de
Bras\'{\i}lia, 70910-900, Bras\'\i lia, DF, Brazil}

\author{R.G.G. Amorim} \email{ronniamorim@gmail.com}
\affiliation{International Center for Condensed Matter Physics, Instituto de F\'{\i}sica, Universidade de
Bras\'{\i}lia, 70910-900, Bras\'\i lia, DF, Brazil}
\affiliation{Faculdade
Gama, Universidade de Bras\'{\i}lia, 72444-240, Bras\'{\i}lia, DF,
Brazil.}

\author{S. C. Ulhoa}\email{sc.ulhoa@gmail.com}
\affiliation{International Center for Condensed Matter Physics, Instituto de F\'{\i}sica, Universidade de
Bras\'{\i}lia, 70910-900, Bras\'\i lia, DF, Brazil}

\author{A. E. Santana}\email{a.berti.santana@gmail.com}
\affiliation{International Center for Condensed Matter Physics, Instituto de F\'{\i}sica, Universidade de
Bras\'{\i}lia, 70910-900, Bras\'\i lia, DF, Brazil}

\begin{abstract}
The Schr\"{o}dinger equation in phase space is used to calculate the Wigner function for the Helium atom in the approximation of a system of two oscillators. Dissipation effect is analysed and the non-classicality of the state is studied by  the non-classicality indicator of the Wigner function, which is calculated as a function of the dissipation parameter.
\end{abstract}


\maketitle

\section{Introduction}
Dissipative systems in quantum mechanics have raised interest for
many years~\cite{d1,d2}. A particular reason is that such systems
are temporally irreversible; i.e.  there is a preferred time
direction in their evolution. Irreversible phenomena, such as
dissipation, emerges out of interaction of a system with its
neighborhoods,  such that the energy flows
irreversibly~\cite{d3,d4}. Thus for instance a dissipative force
as friction could find a microscopic explanation. In a broad
perspective, many microscopic phenomena are described by
irreversible models~\cite{d5,d6,d7,d8,d9}. It is important to
emphasize that the study of irreversible systems in such a
framework can shed some light into the very structure of matter,
in particular in experimental apparatus with atoms where the
neighborhood effect  is the cause of dissipation.

This is the case of  some procedures with fermions following in parallel
with the production and detection of Bose-Einstein
condensates, using the expedients such as laser cooling, magnetic and
magneto-optic traps~\cite{fer1,fer2}. The fermion counterpart has been
accomplished by considering a degenerate Fermi gas as well as condensates of
rare isotopes ~\cite{fer2,fer3,fer4,fer5,fer6}. In addition, there is a
great deal of interest in studying entangled multipartite fermion states for
quantum communication~\cite{fer12,fer121}.  A simple but intricate
example of such a fermion system, at the level of electronic structure, is the
Helium atom considered as a few-fermion system taken in a external field, which can in turn be considered as a dissipative effect.

For practical analysis, it is interesting to note that a combination of two damped quantum harmonic oscillators can describe an atom with two electrons. In this case, the Schr\"{o}dinger equation has no known exact solution. The results obtained for
such systems are based on approximative methods or  variational
formalisms. However, due to the resemblance between the gaussian wave function of the spherically symmetric harmonic oscillator and the 1$s$ state
of the hydrogen atom, some models are used to study solutions of the Schr\"{o}dinger equation for Helium atom. It  consists in changing the Coulomb interactions by the harmonic oscillator potential. In particular, in
the work of Kestner~\cite{he1,he2}, the electron-nuclei interactions were
replaced by a harmonic oscillator potential but the electron-electron interaction was Coulombic. Then it was shown that the energy values obtained were very close to experimental data. In the presented work we address this problem, in order to study the non-classicality of such states, considering in addition,
dissipation effects. We proceed by using the quantum mechanics in phase space in order to analyse the Wigner function~\cite{wig1,wig2,wig3,wig4,2gal1}.

The analysis in phase space is important in order to track evidences of chaoticity as well as
the statistical nature of quantum states. In this case, the Wigner formalism is physically appealing, in particular in experiments for the reconstruction of quantum states, in  quantum tomography and  for the direct measurement of the Wigner function~\cite%
{smithey,leibfried,davidovich,vogel,Ibort:2009bk,Ibort:2010xz,Ibort:2012ab}.

However, there are difficulties with the direct use of the Wigner function. One is that, there is no gauge symmetry associated with the Wigner function, since it is a real function. This aspect is also a problem for considering superposition effects in phase space. Another difficulty is related with technical reasons: the Liouville-von Neumann equation in phase space has an intricate nature, with no practical perturbative solution. For
instance, a perturbative theory for fermions is a non-trivial task and
remains to be formulated consistently. This type of problem has lead to an intense search for the analysis of the Wigner formalism~\cite{1dodonov1,dodo7,dodo8,olavo3,olavo4,boo1,seb3,seb222,%
seb8,seb9,seb10,seb12,seb13}, and some advancements have been made; it includes the study of representations of quantum equations directly in
phase-space~\cite{seb42}. After some preliminary attempts~\cite{vega1,vega2}, exploring a representation of the Schr\"{o}dinger equation in phase space~\cite{dayi,gosson1,gosson2,gosson3,gosson4}, a consistent formalism has been introduced~\cite{2kha2}, by using the notion of quasi-amplitude of probabilities, which is associated with the Wigner function by the Moyal (or star) product in phase space~\cite{moy1,moy2}. This
notion of symplectic structure and Weyl product have been explored to study unitary representations of Galilei group, leading to a symplectic (phase space) representation of Schr\"{o}dinger equation~\cite{2kha2}. This approach provides an interesting procedure to deriving the
Wigner function, by using consistently the gauge invariance~\cite{gauge1}. This symplectic representation was  applied in kinetic theory and extended to the relativistic context; giving rise to the Klein-Gordon and the Dirac
equations in phase space~\cite{seb2, seb22, ron1, ron2,sig1}. Here we use this symplectic quantum mechanics to analyse the behaviour of the Wigner function for the Helium atom, considering dissipation. In this case, we study the non-classicality of the state  with the non-classicality (negativity) indicator  of the Wigner function~\cite{kenfack1} as a  function of the dissipation parameter.

The paper is organized in the following way. In Section II,  we present an outline of the symplectic representation
the Schr\"{o}dinger equation in phase space and the connection between phase space quasi-amplitudes and the Wigner function. In Section III, we solve the Schr\"{o}dinger equation in phase space for the Helium atom in the two-oscillator approximation. In Section IV, a quantum damped oscillator is studied. Finally,
some closing comments are given in Section V.

\section{Outline on Schr\"{o}dinger equation in phase space}

In this section we present a brief outline of the construction  of the Schr\"{o}dinger equation in
phase space, emphasizing the association of phase space amplitude of probability with the Wigner function. We consider initially a one-particle
system described by the Hamiltonian $H=\widehat{p}^{2}/2m$, where $m$ and $%
\widehat{p}$ are the mass and the momentum, respectively, of the particle.
The Wigner formalism for such a system is constructed from the Liouville-von
Neumann equation~\cite{wig1, wig2, wig3, wig4}%
\begin{equation*}
i\hbar\partial_{t}\rho(t)=[H,\rho],
\end{equation*}
where $\rho(t)\;$\ is the density matrix. The Wigner function, $f_{w}(q,p),$
is defined by
\begin{equation}
f_{W}(q,p)=(2\pi\hbar)^{-3}\int dz\exp(\frac{ipz}{\hbar})\langle q-\frac{z}{2%
}|\rho|q+\frac{z}{2}\rangle,   \label{wigago51}
\end{equation}
and satisfies the equation of motion%
\begin{equation}
i\hbar\partial_{t}f_{W}(q,p,t)=\{H_{W},f_{W}\}_{M},   \label{wigago52}
\end{equation}
where $H_W$ is the Wigner Hamiltonian and $\{a,b\}_{M}=a\star b-b\star a$ is the Moyal bracket, such that the
star-product $a\star b$ is given by%
\begin{equation*}
a\star b=a(q,p)e^{\frac{i\hbar\Lambda}{2}}b(q,p)
\end{equation*}
with $\Lambda=\overleftarrow{\partial}_{p}\overrightarrow{\partial}_{q}-%
\overleftarrow{\partial}_{q}\overrightarrow{\partial}_{p}.$ The functions $%
a(q,p)$ are defined in a manifold $\Gamma$, using the basis ($q,p$) with the
physical content of the phase space. In this formalism an operator, say $A,$
defined in the Hilbert space $\mathcal{H}$, is represented by the function
\begin{equation*}
A(q,p)=\int dz\exp(\frac{ipz}{\hbar})\langle q-\frac{z}{2}|A|q+\frac{z}{2}%
\rangle,
\end{equation*}
such that the product of two operators, $AB$, reads%
\begin{equation*}
(AB)(q,p)=A(q,p)e^{\frac{i\hbar\Lambda}{2}}B(q,p)=A(q,p)\star B(q,p).
\end{equation*}
The average of the operator $A$ in a state $\psi\in\mathcal{H}$ is given by%
\begin{equation*}
\langle A\rangle=\langle\psi|A|\psi\rangle=\int dqdpA(q,p)f_{W}(q,p)=Tr\rho
A.
\end{equation*}

Now we proceed in order to introduce the symplectic representation of quantum mechanics in phase space. First we  introduce a Hilbert space associated to the phase space $\Gamma$,  by considering the set of function $\phi(q,p)$ in $\Gamma$, such that
\begin{equation*}
\int dpdq\psi^{\ast}(q,p)\psi(q,p)<\infty\
\end{equation*}
is a bilinear real form. This Hilbert space is denoted by $\mathcal{H}%
(\Gamma)$. Unitary mappings, $U(\alpha)$, in $\mathcal{H}(\Gamma)$ are
naturally introduced by using the star-product, i.e. \
\begin{equation*}
U(\alpha)=\exp(\alpha\widehat{A}),
\end{equation*}
where
\begin{align*}
\widehat{A} & =A(q,p)\star=A(q,p)\exp\left[ \frac{i\hbar}{2}(\frac {%
\overleftarrow{\partial}}{\partial q}\frac{\overrightarrow{\partial}}{%
\partial p}-\frac{\overleftarrow{\partial}}{\partial p}\frac{%
\overrightarrow {\partial}}{\partial q})\right] \\
& =A(q+\frac{i\hbar}{2}\partial_{p},p-\frac{i\hbar}{2}\partial_{p}).
\end{align*}
Let us consider some examples. For the basic functions $q$ and $p$
(3-dimensional Euclidian vectors), we have
\begin{equation}
\widehat{q}_{i}=q_{i}\star=q_{i}+\frac{i\hbar}{2}\partial_{p_{i}},
\label{eq 13}
\end{equation}%
\begin{equation}
\widehat{p}_{i}=p_{i}\star=p_{i}-\frac{i\hbar}{2}\partial_{q_{i}}.
\label{eq 14}
\end{equation}
These operators satisfy the Heisenberg relations $\left[ \widehat{q}_{j},%
\widehat{p}_{l}\right] =i\hbar\delta_{jl}$. Then we introduce a Galilei
boost by defining the boost generator $\widehat{k}_{i}=mq_{i}\star-tp_{i}%
\star=m\widehat{q}_{i}-t\widehat{p}_{i}$, $i=1,2,3$, such that
\begin{align*}
\exp\left( -i\mathbf{v}\cdot\widehat{\mathbf{k}}/\mathbb{\hbar}\right)
\widehat{q}_{j}\exp\left( i\mathbf{v}\cdot\widehat{\mathbf{k}}/\mathbb{\hbar
}\right) & =\widehat{q}_{j}+v_{j}t\mathbf{,} \\
\exp\left( -i\mathbf{v}\cdot\widehat{\mathbf{k}}/\mathbb{\hbar}\right)
\widehat{p}_{j}\exp\left( i\mathbf{v}\cdot\widehat{\mathbf{k}}/\mathbb{\hbar
}\right) & =\widehat{p}_{j}+mv_{j}\mathbf{.}
\end{align*}
These results, with the commutation relations, show that $\widehat{q}$ and $%
\widehat{p}$ are physically the position and momentum operators,
respectively.

We introduce the operators $\overline{Q}$ and $\overline{P}$, such that $[%
\overline{Q},\overline{P}]=0$, $\ \overline{Q}|q,p\rangle=q|q,p\rangle$ and $%
\overline{P}|q,p\rangle=p|q,p\rangle$, with
\begin{equation}
\langle q,p|q^{\prime},p^{\prime}\rangle=\delta(q-q^{\prime})\delta
(p-p^{\prime}),  \notag
\end{equation}
and $\int dqdp|q,p\rangle\langle q,p|=1$. From a physical point of view, we
observe the transformation rules:
\begin{equation*}
\exp(-iv\frac{\widehat{k}}{\hbar})2\overline{Q}\exp(iv\frac{\widehat{k}}{%
\hbar})=2\overline{Q}+vt\mathbf{1},
\end{equation*}
and
\begin{equation*}
\exp(-iv\frac{\widehat{k}}{\hbar})2\overline{P}\exp(iv\frac{\widehat{k}}{%
\hbar})=2\overline{P}+mv\mathbf{1}.
\end{equation*}
Then $\overline{Q}$ and $\overline{P}$ are transformed, under the Galilei
boost, as position and momentum, respectively. Therefore, the manifold
defined by the set of eigenvalues $(q,p)$ has the content of a phase
space. However, the operators $\overline{Q}$ and $\overline{P}$ are not
observables, since they commute with each other.

Considering a homogeneous systems satisfying the Galilei symmetry, the commutations relation between $\widehat{k}$ and $\widehat{H}$ is $[%
\widehat{k}_{j},\widehat{H}]=i\widehat{P}_{j}$. Explicitly, we  have%
\begin{equation*}
\lbrack mq_{j}+i\hbar\frac{\partial}{\partial p_{j}},H(q,p)\star]=ip_{j}+\frac {\hbar%
}{2}\frac{\partial}{\partial q_{j}}.
\end{equation*}
A solution, providing a general form to $\widehat{H}=H(q,p)\star$, is
\begin{align}
\widehat{H} & =\frac{p^{2}\star}{2m}+V(q)\star  \notag \\
& ={\frac{p^{2}}{2m}}-{\frac{\hbar^{2}}{8m}}{\frac{\partial^{2}}{\partial
q^{2}}}-{\frac{i\hbar p}{2m}}{\frac{\partial}{\partial q}}+V(q\star) {.}
\label{ago20.2}
\end{align}
This is the Hamiltonian of a one-body system in an external field.

Consider the time evolution of a state $\psi(q,p;t)$, that is given by $%
\psi(q,p;t)=U(t,t_{0})\psi(q,p;t_{0}),$ where $U(t,t_{0})=\exp(-i\hbar
(t-t_{0})\widehat{H})$. This result leads to a Schr\"{o}dinger-like equation
written in phase-space, i.e. \cite{2kha2}%
\begin{equation}
i\hbar\partial_{t}\psi(q,p,t)=\widehat{H}\psi(q,p,t)   \label{ago25.1}
\end{equation}

Now physical
meaning of the state $\psi(q,p,t)$ has to be identified. This is done, by associating $%
\psi(q,p,t) $ with the Wigner function. From Eq.~(\ref{ago25.1}), one can prove that $g(q,p)=\psi(q,p,t)\star\psi^{\dagger}(q,p,t)$ satisfies Eq.~(\ref{wigago52})~\cite{2kha2,seb2,gauge1}. In addition, using the associative
property of the Moyal product and the relation
\begin{equation*}
\int dqdp\psi(q,p,t)\star\psi^{\dagger}(q,p,t)=\int dqdp\psi(q,p,t)\psi
^{\dagger}(q,p,t),
\end{equation*}
we have
\begin{align*}
\langle A\rangle & =\langle\psi|A|\psi\rangle \\
& =\int dqdp\psi(q,p,t)\widehat{A}(q,p)\psi^{\dagger}(q,p,t) \\
& =\int dqdpf_{W}(q,p,t)A(q,p,t),
\end{align*}
where $\widehat{A}(q,p)=A(q,p)\star$ \ is an observable. Thus, the Wigner function can be calculated by using
\begin{equation}  \label{wig1}
f_{W}(q,p)=\psi(q,p)\star\psi^{\dagger}(q,p).
\end{equation}
It is to be noted also that the eigenvalue equation,
\begin{equation}
H(q,p)\star\psi=E\psi,
\end{equation}
results in $H(q,p)\star f_{W}=Ef_{W}.$ Therefore, $\psi(q,p)$ and $%
f_{W}(q,p) $ satisfy the same differential equation. These results show that
Eq.~(\ref{ago25.1}) is a fundamental starting point for the description of
quantum physics in phase space, fully compatible with the Wigner formalism.

\section{Helium-like system in phase space}

Although Schr\"{o}dinger equation can not be accurately solved, there are Helium-like systems that admit exact solutions. In this section we consider
the Helium-like system , such that the Coulomb interaction is replaced by Hooke-like forces, including electron-electron interaction~\cite{he1,he2,he3}.Then we uses the Schr\"{o}dinger equation in phase space to
obtain the Wigner function for the Helium-like atom in this approximation. The classical counterpart of the Helium-like atom Hamiltonian  is then written as~\cite{he3}
\begin{equation}
H=\frac{p_{1}^{2}}{2m}+\frac{p_{2}^{2}}{2m}+\frac{1}{2}m\omega
^{2}(x_{1}^{2}+x_{2}^{2})-\frac{\xi }{4}(x_{1}-x_{2})^{2},  \label{1}
\end{equation}%
where the the sub-indices $1$ and $2$ refer to the electrons and $\xi$ is a small parameter. We restrict
our analysis to the one-dimensional case. Using the variables,
\begin{eqnarray*}
u &=&\frac{x_{1}+x_{2}}{\sqrt{2}},\\v&=&\frac{x_{1}-x_{2}}{\sqrt{2}}, \\
p_{u} &=&\frac{p_{1}+p_{2}}{\sqrt{2}},\\p_{v}&=&\frac{p_{1}-p_{2}}{\sqrt{2}},
\end{eqnarray*}%
Eq.~(\ref{1}) is written as
\begin{equation}
H=\frac{p_{u}^{2}}{2m}+m\omega ^{2}u^{2}+\frac{p_{v}^{2}}{2m}+\frac{%
(1-\xi )}{2}m\omega ^{2}v^{2}.  \label{hamhe}
\end{equation}%
It is convenient to write  $H=H_{u}+H_{v}$, where%
\begin{equation*}
H_{u}=\frac{p_{u}^{2}}{2m}+m\omega ^{2}u^{2},
\end{equation*}%
and
\begin{equation*}
H_{v}=\frac{p_{v}^{2}}{2m}+\frac{(1-\xi )}{2}m\omega ^{2}v^{2}.
\end{equation*}

The time-independent Schr\"{o}dinger equation in phase space is written as
\begin{equation}  \label{4}
H\star\psi(u,v,p_{u},p_{v})=E\psi(u,v,p_{u},p_{v}).
\end{equation}
In order to solve this equation, we take
\begin{equation*}
\psi (u,v,p_{u},p_{v})=\varphi (u,p_{u})\chi (v,p_{v}),
\end{equation*}%
and
\begin{equation*}
E=E_{u}+E_{v}.
\end{equation*}%
It is important to consider the relations,
\begin{eqnarray*}
u\star &=&u+\frac{i\hbar }{2}\frac{\partial }{\partial p_{u}},\\p_{u}\star
&=&p_{u}-\frac{i\hbar }{2}\frac{\partial }{\partial u},\\
v\star &=&v+\frac{i\hbar }{2}\frac{\partial }{\partial p_{v}},\\p_{v}\star
&=&p_{v}-\frac{i\hbar }{2}\frac{\partial }{\partial v},
\end{eqnarray*}%
which are obtained from the star product in Eq.~(\ref{4}). In this sense, the resulting
equations are solved by starting from the equation for the coordinates $u$ and $p_{u}$; i.e.
\begin{equation}
\left( \frac{p_{u}^{2}\star }{2m}+m\omega ^{2}u^{2}\star \right) \varphi
_{n}=E_{u}\varphi _{n}.  \label{5}
\end{equation}%
Writing
\begin{equation}
H_{u}\star =\frac{m\omega ^{2}}{2}\left( u\star +\frac{i}{m\omega }%
p_{u}\star \right) \left( u\star -\frac{i}{m\omega }p_{u}\star \right)
-\hbar \omega,  \label{6}
\end{equation}%
we then introduce the operators
\begin{equation}
a_{u}\star =\sqrt{\frac{m\omega }{2\hbar }}\left( u\star +\frac{i}{m\omega }%
p_{u}\star \right) ,  \label{7}
\end{equation}%
\begin{equation}
a_{u}^{\dagger }\star =\sqrt{\frac{m\omega }{2\hbar }}\left( u\star -\frac{i%
}{m\omega }p_{u}\star \right),  \label{8}
\end{equation}%
satisfying  the relations,
\begin{eqnarray*}
\lbrack a_{u}\star ,a_{u}^{\dagger }\star ] &=&1, \\
a_{u}\star \varphi _{n} &\propto &\varphi _{n-1},\\a_{u}^{\dagger }\star
\varphi _{n}&\propto& \varphi _{n+1},
\end{eqnarray*}%
where $n=0,1,2,\ldots $, and $a_{u}\star \varphi _{0}=0$.

The
Hamiltonian given in Eq.~(\ref{6}) is then written as
\begin{equation}
H_{u}\star =\hbar \omega (a_{u}^{\dagger }\star a_{u}\star +\frac{1}{2}).
\label{9}
\end{equation}%
In this way, we have
\begin{equation}
\sqrt{\frac{m\omega }{2\hbar }}\left( u\star +\frac{i}{m\omega }p_{u}\star
\right) \varphi _{0}=0.\label{9b}
\end{equation}%
Substituting $u\star =u+\frac{i\hbar }{2}\frac{\partial }{\partial _{p_{u}}}$ and $p_{u}\star
=p_{u}-\frac{i\hbar }{2}\frac{\partial }{\partial _{u}}$ in Eq.(\ref{9b}) we obtain
\begin{equation}
\sqrt{\frac{m\omega }{2\hbar }}\left[u+\frac{i\hbar }{2}\frac{\partial }{\partial _{p_{u}}}+\frac{i}{m\omega }\left(p_{u}-\frac{i\hbar }{2}\frac{\partial }{\partial _{u}}\right)
\right] \varphi _{0}=0.\label{9c}
\end{equation}%
Separating the real and imaginary part of Eq.~(\ref{9c}), and considering $\varphi_{0}(u,p_{u})=\varphi^{a}_{0}(u)\varphi^{b}_{0}(p_{u})$, we can show that real part satisfies the
differential equation
\begin{equation}
u\varphi _{0}+\frac{\hbar }{2m\omega }\frac{\partial \varphi^{a} _{0}}{\partial u%
}=0,  \label{10}
\end{equation}%
with a solution given by
\begin{equation}
\varphi^{a} _{0}(u)=\exp \left( -\frac{2m\omega }{\hbar }u^{2}\right) .  \label{11}
\end{equation}%
For the imaginary part, we have
\begin{equation}
\frac{\hbar }{2}\frac{\partial \varphi _{0}}{\partial p_{u}}+\frac{p_{u}}{%
m\omega }\varphi^{b} _{0}=0,  \label{12}
\end{equation}%
with the solution
\begin{equation}
\varphi^{b} _{0}(p_{u})=\exp \left( -\frac{2}{\hbar m\omega }p_{u}^{2}\right) .
\label{13}
\end{equation}%
Then, we get
\begin{equation}
\varphi _{0}(u,p_{u})\sim \exp \left( -\frac{2m\omega }{\hbar }u^{2}-\frac{2%
}{\hbar m\omega }p_{u}^{2}\right) .  \label{14}
\end{equation}

Similarly, the solution of the equation for $\chi $ is obtained, i.e.
\begin{equation}
\left( \frac{p_{v}^{2}\star }{2m}+\frac{(1-\xi )}{2}m\omega
^{2}v^{2}\star \right) \chi _{n}=E_{v}\chi _{n}.  \label{15}
\end{equation}%
This leads to
\begin{eqnarray}
H_{v}\star =&&\frac{(1-\xi m\omega ^{2})}{2}\left( v\star +\frac{i}{%
m\omega (1-\xi)^{1/2}}p_{v}\star \right)   \notag \\
&\times &\left( v\star -\frac{i}{m\omega (1-\xi)^{1/2}}p_{v}\star
\right) -\frac{\hbar \omega }{2}(1-\xi)^{1/2}.  \label{16}
\end{eqnarray}%
Then, we define
\begin{equation}
a_{v}\star =\sqrt{\frac{m\omega }{2\hbar }}\left( v\star +\frac{i}{m\omega
(1-\xi)^{1/2}}p_{v}\star \right) ,  \label{17}
\end{equation}%
and
\begin{equation}
a_{v}^{\dagger }\star =\sqrt{\frac{m\omega }{2\hbar }}\left( v\star -\frac{i%
}{m\omega (1-\xi)^{1/2}}p_{v}\star \right) .  \label{18}
\end{equation}%
These operators satisfy the relations,
\begin{eqnarray*}
[a_{v}\star ,a_{v}^{\dagger
}\star ]&=&1,\\a_{v}\star \chi _{n}&\propto& \chi _{n-1},\\ a_{v}^{\dagger
}\star \chi _{n}&\propto& \chi _{n+1},
\end{eqnarray*}
where $n=0,1,2,\ldots $.

The operator
given in Eq.~(\ref{16}) has the form
\begin{equation}
H_{v}\star =\hbar \omega (a_{v}^{\dagger }\star a_{v}\star -\frac{(1-\xi)^{1/2}}{2}).  \label{19}
\end{equation}%
We can show that
 $a_{v}\star \chi _{0}=0$, such that
\begin{equation}
\sqrt{\frac{m\omega (1-\xi)}{2\hbar }}\left( v\star +\frac{i}{m\omega
(1-\xi)^{1/2}}p_{v}\star \right) \chi _{0}=0. \label{19b}
\end{equation}%

The real and imaginary part of Eq.~(\ref{19b})  leads,
respectively, to the solutions
\begin{equation}
\chi _{0}=\exp \left( -\frac{2m\omega (1-\xi)^{1/2}}{\hbar }%
v^{2}\right)    \label{20}
\end{equation}
and
\begin{equation}
\chi _{0}=\exp \left( -\frac{2}{\hbar m\omega (1-\xi)^{1/2}}%
p_{v}^{2}\right) ,  \label{21}
\end{equation}%
such that
\begin{equation}
\chi _{0}(v,p_{v})\sim \exp \left( -\frac{2m\omega (1-\xi)^{1/2}}{\hbar
}v^{2}-\frac{2}{\hbar m\omega (1-\xi)^{1/2}}p_{v}^{2}\right) .
\label{22}
\end{equation}

Therefore, the zero order solution of the Schr\"{o}dinger equation is
\begin{eqnarray}
\psi _{0}(u,v,p_{u},p_{v}) &=&  \frac{2e}{\pi \hbar } \exp( -\frac{2m\omega }{\hbar }[u^{2}+(1-\xi)^{1/2}v^{2}])
\notag \\
&\times &\exp ( -\frac{2}{m\omega \hbar }[p_{u}^{2}+(1-\xi
)^{-1/2}p_{v}^{2}]), \notag 
\end{eqnarray}%
where we have used the normalization condition
\begin{equation*}
\int dudvdp_{u}dp_{v}\psi _{n}^{\dagger }(u,v,p_{u},p_{v})\star \psi
_{n}(u,v,p_{u},p_{v})=1.
\end{equation*}

To obtain higher order wave functions, we use the relation
\begin{equation}
\psi _{n}(u,v,p_{u},p_{v})=(a_{u}^{\dagger }\star a_{v}^{\dagger }\star
)^{n}\psi _{0}(u,v,p_{u},p_{v}).  \label{24}
\end{equation}%

The Wigner function is found from
\begin{equation}
f_{W}^{(n)}(u,v,p_{u},p_{v})=\psi _{n}(u,v,p_{u},p_{v})\star \psi
_{n}^{\dagger }(u,v,p_{u},p_{v}). \notag 
\end{equation}%
In particular for $n=0$ we obtain
\begin{eqnarray}
f_{W}^{(0)}(q_{1},q_{2},p_{1},p_{2}) &=&\left( \frac{2e}{\pi \hbar }\right)
\exp \left( -\frac{2m\omega }{\hbar }\frac{(x_{1}+x_{2})^{2}}{2}\right)
\notag \\
&\times &\exp \left( -\frac{2m\omega }{\hbar }(1-\xi)^{1/2}\frac{%
(x_{1}-x_{2})^{2}}{2}\right)   \notag \\
&\times &\exp \left( -\frac{2}{m\omega \hbar }\frac{(p_{1}+p_{2})^{2}}{2}%
\right)   \notag \\
&\times &\exp \left( -\frac{2}{m\omega \hbar }(1-\xi )^{-1/2}\frac{%
(p_{1}-p_{2})^{2}}{2}\right) .  \notag
\end{eqnarray}%
Hence, the energy of the fundamental state is given by
$E_{0}=\hbar \omega (1-\frac{\xi}{4})$.

These results are interesting in a double sense. First, we have
calculated analytically the Wigner function for Helium-like atom.
Second, the Wigner function has many applications, among them one
stands out quantum computing. So, such a procedure to study the
Wigner function for Helium-like atom opens up new possibilities
for analyzing entanglement. In this context, in experiments, the
dissipation due to the effect of external fields, are a crucial
factor. In the next section, in order to consider the Helium atom
in a non-conservative external field, we add to the Hooke-like
force, a linear dissipation.

\section{Damped Quantum Oscillator}

In order to analyze dissipation effect of neighborhood, here, we
solve the Schr\"{o}dinger equation in phase space for Hooke-like
system with a damped interaction. This stands for the Helium atom
in a dissipative field. We considers a one dimensional system,
where hamiltonian with a dissipative term is (see a different
treatment for such a model in Refs~\cite{d10, d11})
\begin{equation}\label{d1}
\widehat{H}=\frac{1}{2}\left(\widehat{P}^2+\widehat{Q}^2\right)-\frac{\lambda}{2}\left(\widehat{Q}\widehat{P}+\widehat{P}\widehat{Q}\right),
\end{equation}
where $\lambda<1$.

Using the operators given in Eqs.~(\ref{eq 13}) and (\ref{eq 14}) with $\hbar=1$,
\begin{equation}\label{o1}
\widehat{Q}=q+\frac{i}{2}\partial_p,
\end{equation}
and
\begin{equation}\label{o2}
\widehat{P}=p-\frac{i}{2}\partial_q,
\end{equation}
Eq.(\ref{d1}) becomes
\begin{eqnarray}\label{d2}
\widehat{H}&=&\frac{1}{2}\left(p^2+q^2-ip\partial_q+iq\partial_p-\frac{1}{4}\partial^{2}_{q}-\frac{1}{4}\partial^{2}_{p}\right)\nonumber\\
&-&\frac{\lambda}{2}\left(2qp-iq\partial_q+ip\partial_p +\frac{1}{2}\partial_q\partial_p\right).\nonumber
\end{eqnarray}
Applying the this Hamiltonian in the eigenvalue equation $\widehat{H}\psi(q,p)=E\psi(q,p)$, we obtain
\begin{eqnarray}\label{d3}
&&(p^2+q^2)\psi(q,p)-\frac{1}{4}\partial^{2}_{q}\psi(q,p)-\frac{1}{4}\partial^{2}_{p}\psi(q,p)\nonumber\\
&&-\frac{\lambda}{2}\partial_q\partial_p\psi(q,p)-2\lambda qp\psi(q,p)-2E\psi(q,p)=0.\nonumber
\end{eqnarray}

Introducing the new variable
\begin{equation}\label{v1}
z=\frac{1}{2}(p^2+q^2)-\lambda qp,
\end{equation}
we obtain
\begin{equation}\label{d5}
\frac{1}{2}(\lambda^2-1)z\partial^{2}_{z}\psi(z)+ \frac{1}{2}(\lambda^2-1)\partial_{z}\psi(z)+2(z-E)\psi(z)=0.
\end{equation}
Taking $a=\frac{1}{2}(1-\lambda^2)$ and using the ansatz
\begin{equation}\label{d6}
\psi(z)=e^{-\frac{z}{\sqrt{a/2}}}\omega(z),
\end{equation}
we have, after the changing of variables $y=2\sqrt{\frac{2}{a}}z$, the following expression
\begin{equation}\label{d7}
y\partial^{2}_{y}\omega(y)+ (1-y)\partial_{y}\omega(y)-\left[\frac{1}{2}-\frac{E/2}{\sqrt{a/2}}\right]\omega(y)=0.
\end{equation}
The solution of Eq.~(\ref{d7}) is given by the Kummer function (a confluent hypergeometric function, i.e.),
\begin{equation}\label{d8}
\omega(z)=F\left(\frac{1}{2}-\frac{E/2}{\sqrt{a/2}}; 1; 2\sqrt{\frac{2}{a}}z\right).
\end{equation}
In this way, we have the solution
\begin{equation}\label{d9}
\psi(z)=e^{z\frac{1}{\sqrt{a/2}}}F\left(\frac{1}{2}-\frac{E/2}{\sqrt{a/2}}; 1; 2\sqrt{\frac{2}{a}}z\right),
\end{equation}
where $z$ is given in Eq.~(\ref{v1}).
The confluent hypergeometric function condition is such that
$$\frac{1}{2}-\frac{E/2}{\sqrt{a/2}}=-n,$$
where $n \in \mathbb{Z}$.
This relation gives
\begin{equation}\label{d10}
E_{n}=(1-\lambda^2)^{1/2}\left[n+\frac{1}{2}\right].
\end{equation}
Note that if $\lambda=0$ we obtain the result $E_{n}=(n+1/2)$.

The Wigner function can be calculate by
\begin{equation}\label{d11}
f_{W}(q,p,t)=\psi\star\psi^{\ast}.
\end{equation}
In this case we calculate the Wigner functions given in Eq.~(\cite{d11}) using a MAPLE routine. The behavior of the stationary Wigner function for $\lambda=0.1$ are shown in  Figs.~(1)- (4) and for $\lambda=0.9$ are shown in Fig.~(5)-(8).

\begin{figure}[!htb]
\begin{minipage}[b]{\linewidth}
\includegraphics[width=\linewidth]{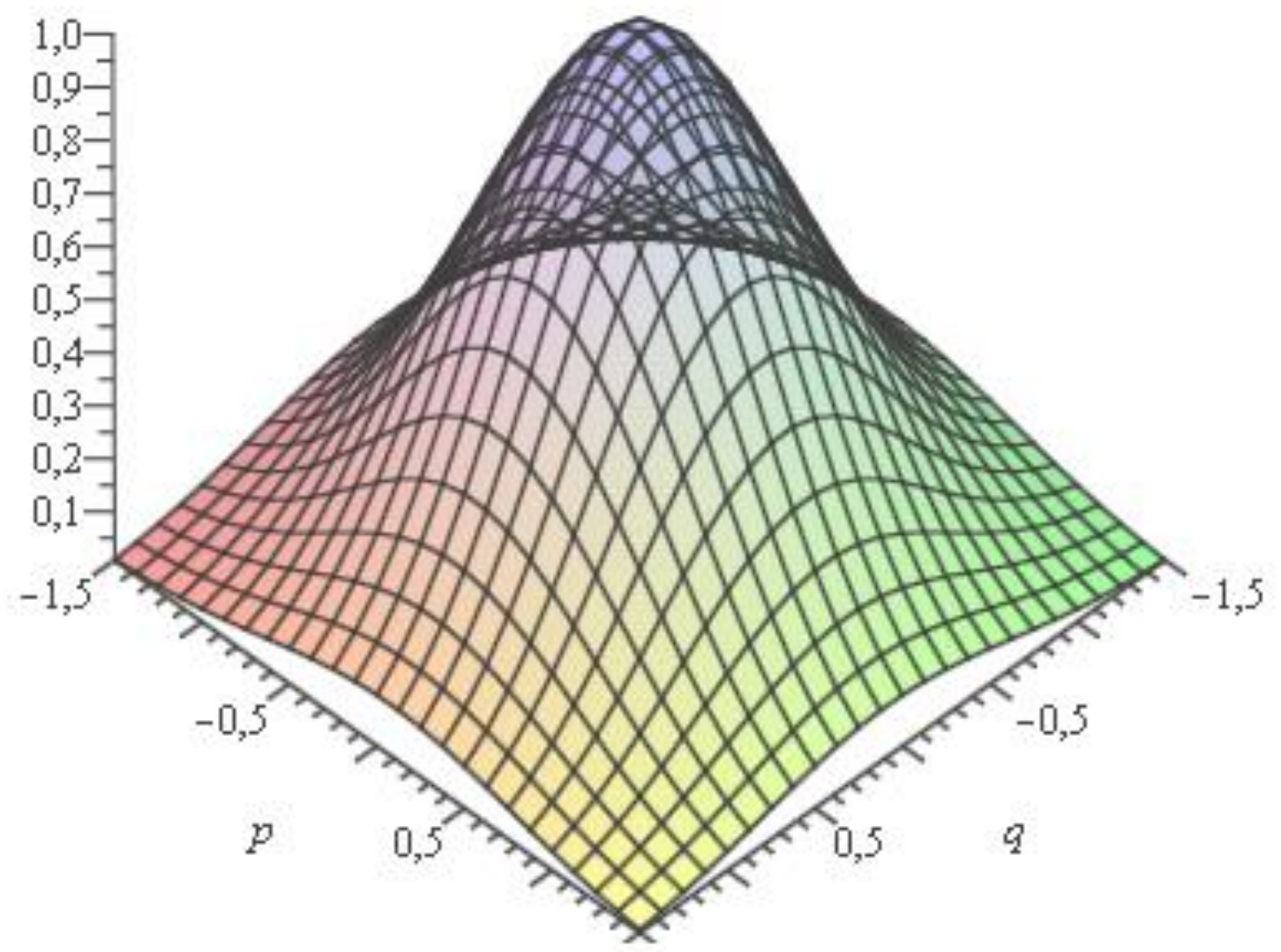}
\caption{Wigner function, $n=0$, $\lambda=0.1$} \label{figura 1}
\end{minipage}\hfill
\begin{minipage}[b]{\linewidth}
\includegraphics[width=\linewidth]{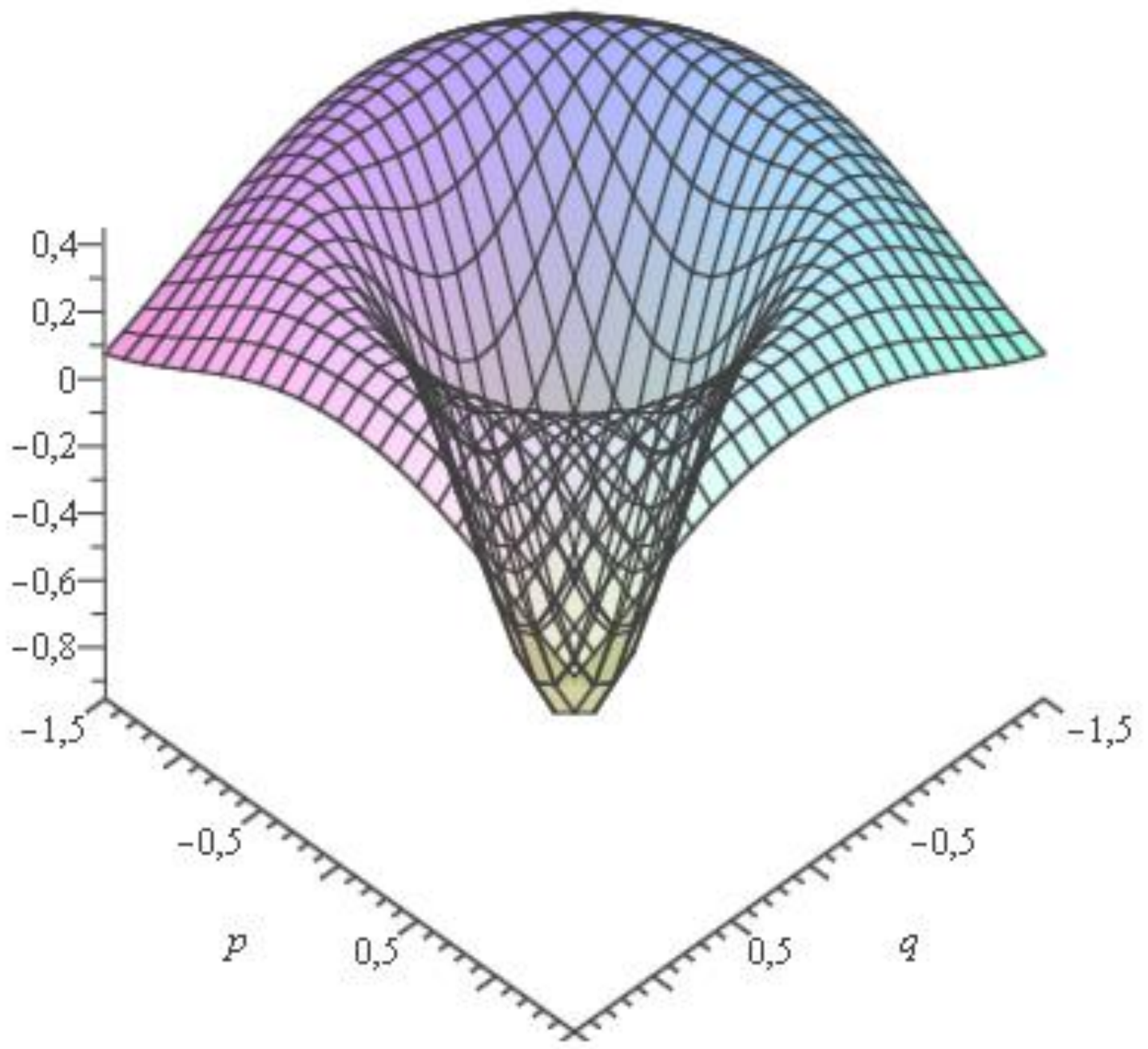}
\caption{Wigner function, $n=1$, $\lambda=0.1$} \label{figura 2}
\end{minipage}
\end{figure}

\begin{figure}[!htb]
\begin{minipage}[b]{\linewidth}
\includegraphics[width=\linewidth]{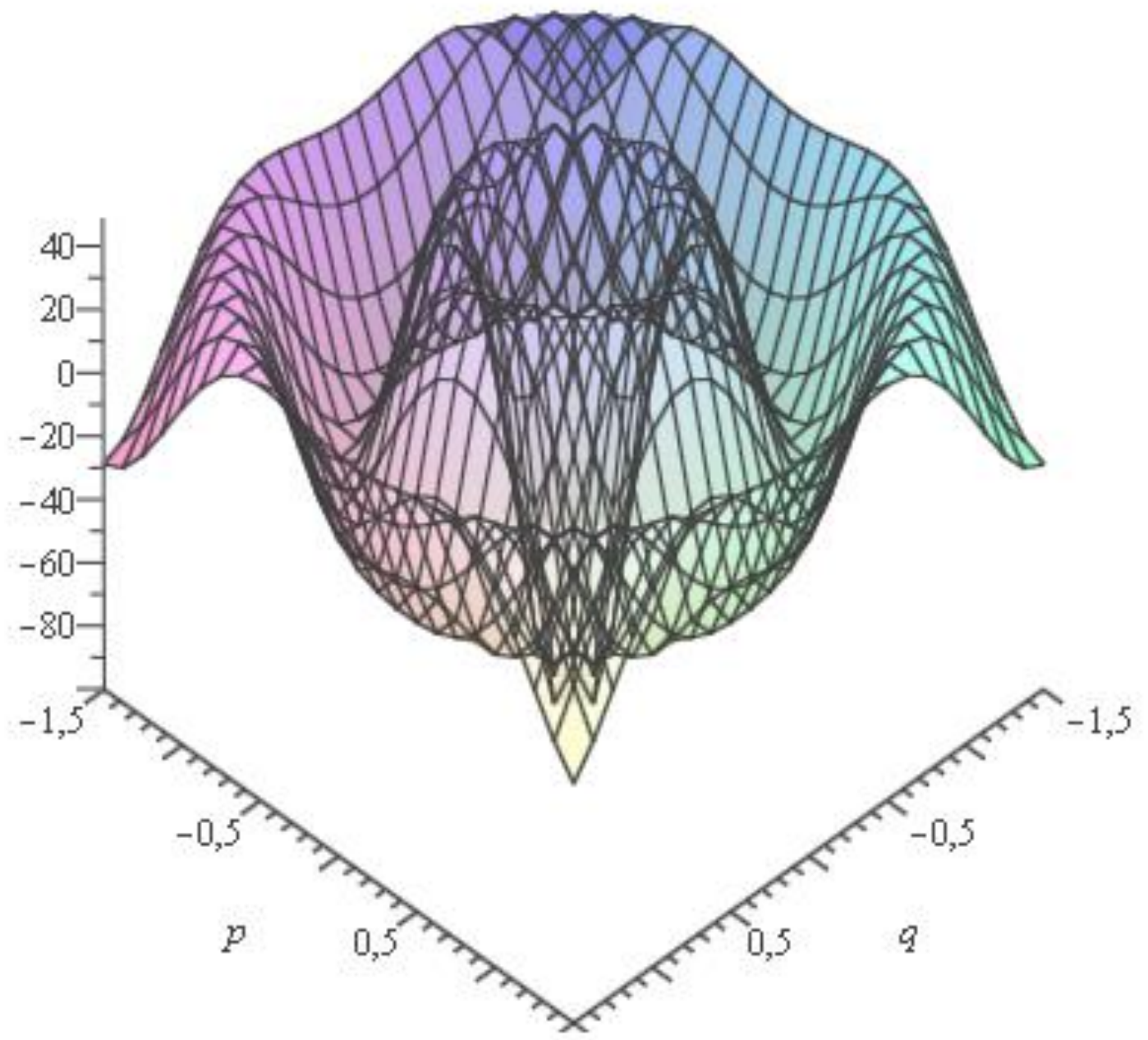}
\caption{Wigner function, $n=5$, $\lambda=0.1$} \label{figura 3}
\end{minipage}\hfill
\begin{minipage}[b]{\linewidth}
\includegraphics[width=\linewidth]{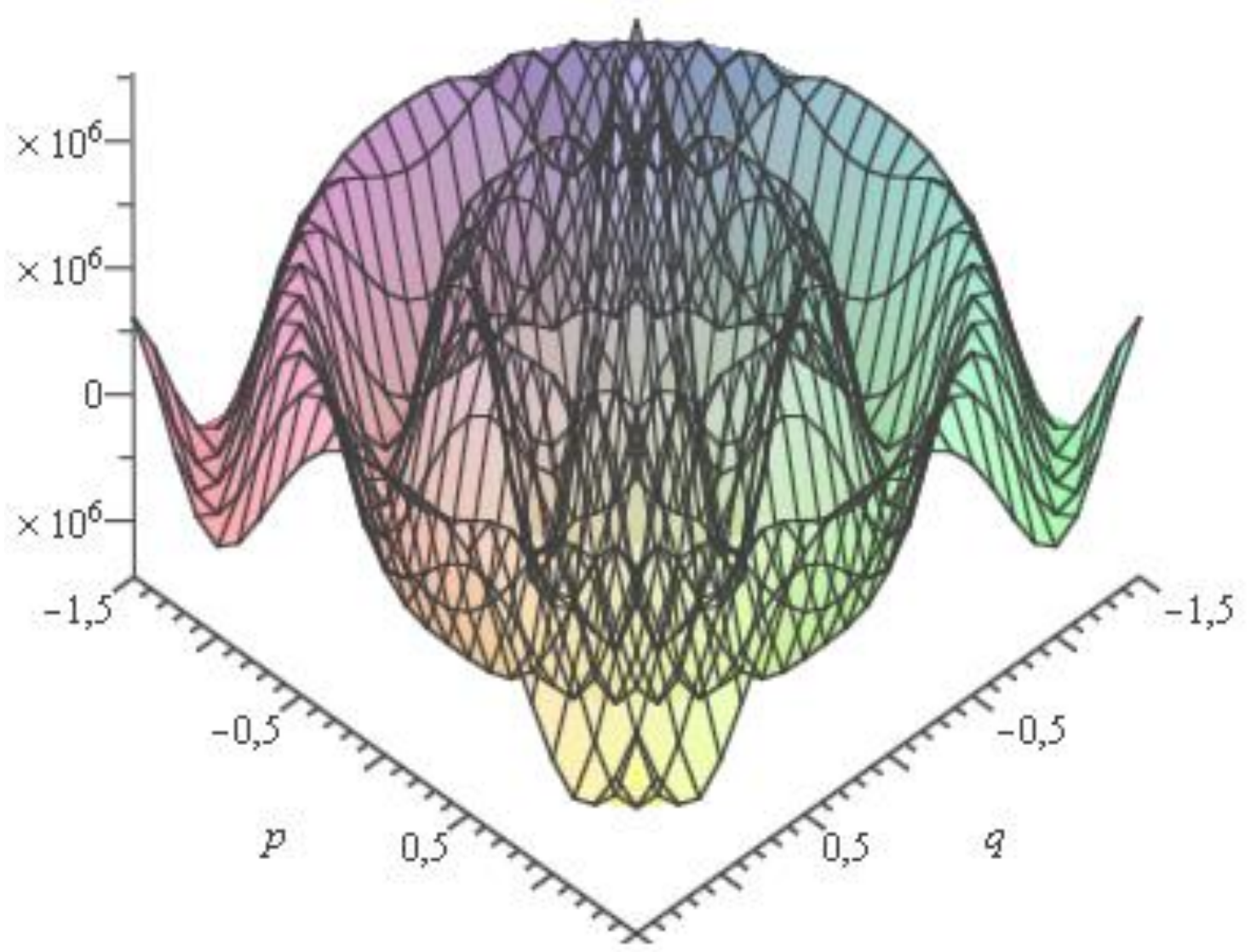}
\caption{Wigner function, $n=10$, $\lambda=0.1$} \label{figura 4}
\end{minipage}
\end{figure}

\begin{figure}[!htb]
\begin{minipage}[b]{\linewidth}
\includegraphics[width=\linewidth]{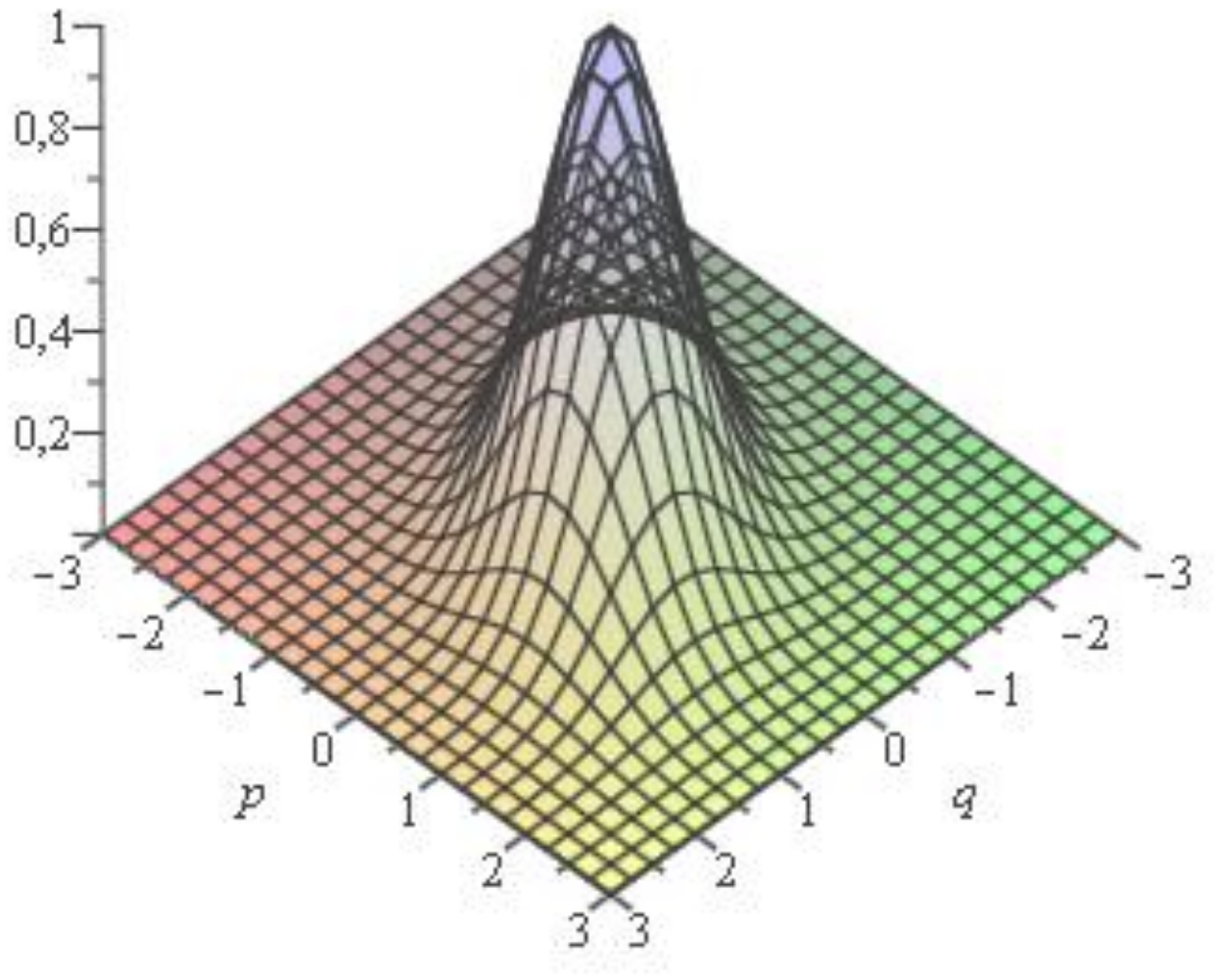}
\caption{Wigner function, $n=0$, $\lambda=0.9$} \label{figura 5}
\end{minipage}\hfill
\begin{minipage}[b]{\linewidth}
\includegraphics[width=\linewidth]{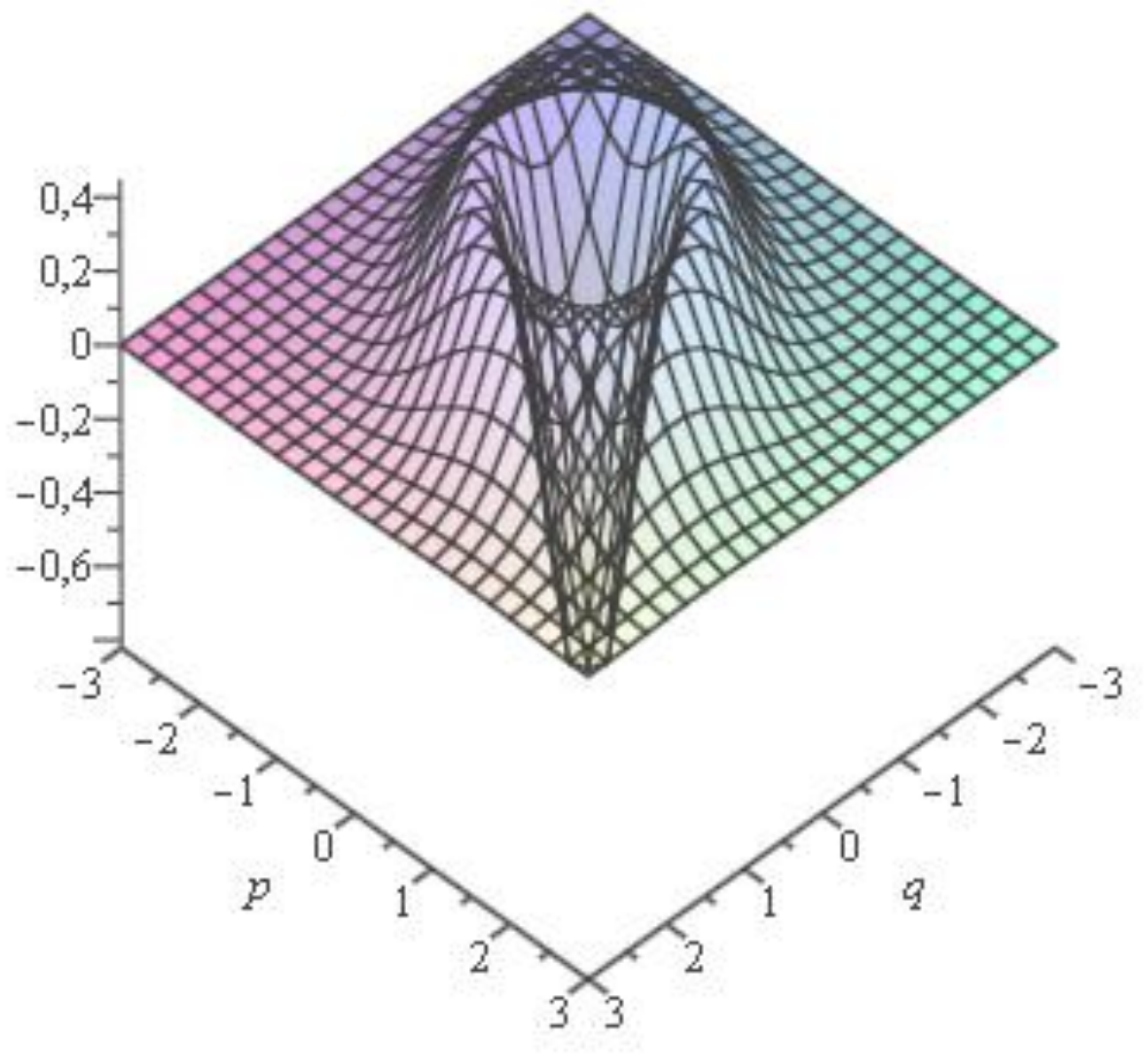}
\caption{Wigner function, $n=1$, $\lambda=0.9$} \label{figura 6}
\end{minipage}
\end{figure}

\begin{figure}[!htb]
\begin{minipage}[b]{\linewidth}
\includegraphics[width=\linewidth]{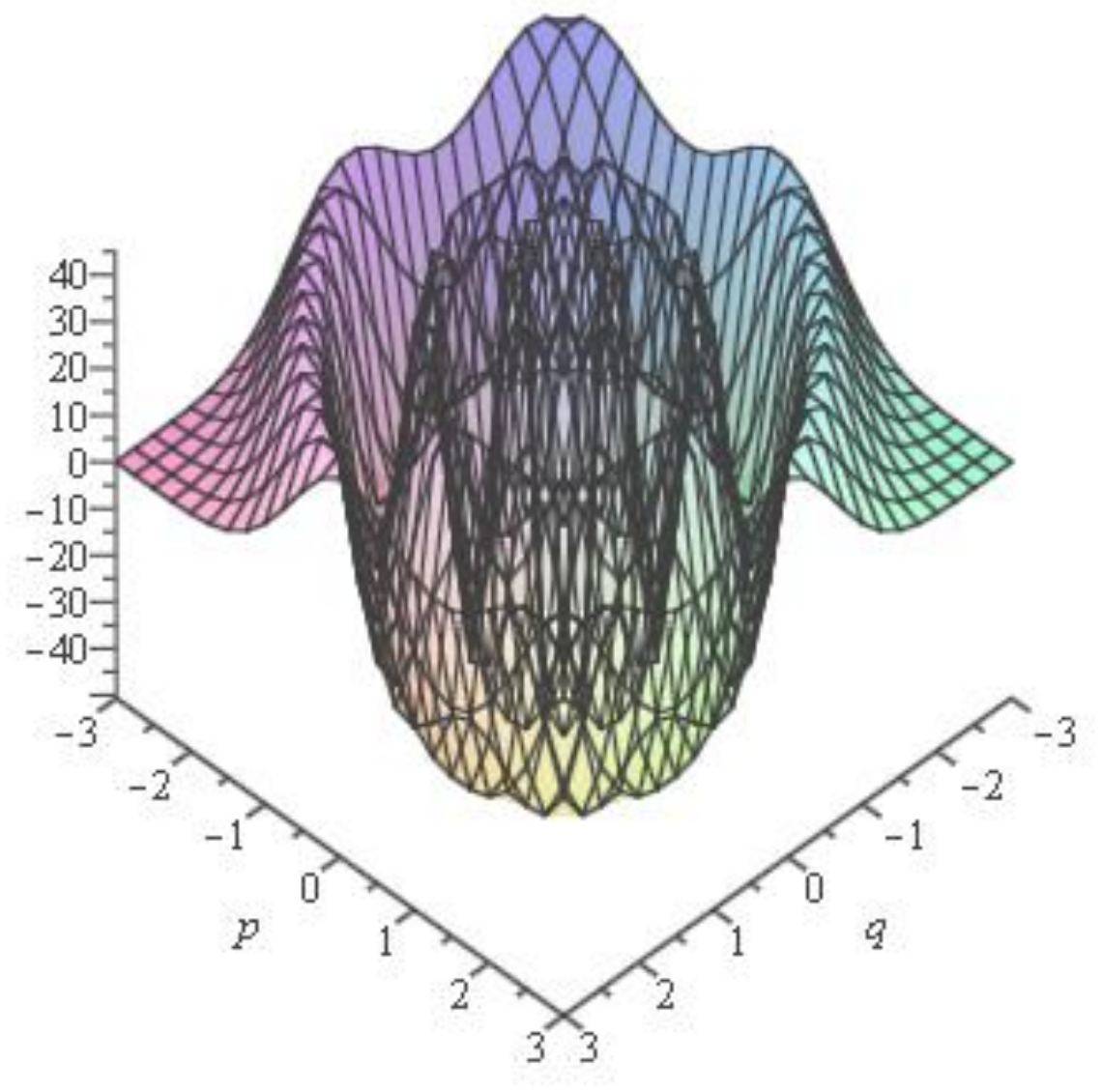}
\caption{Wigner function, $n=5$, $\lambda=0.9$} \label{figura 7}
\end{minipage}\hfill
\begin{minipage}[b]{\linewidth}
\includegraphics[width=\linewidth]{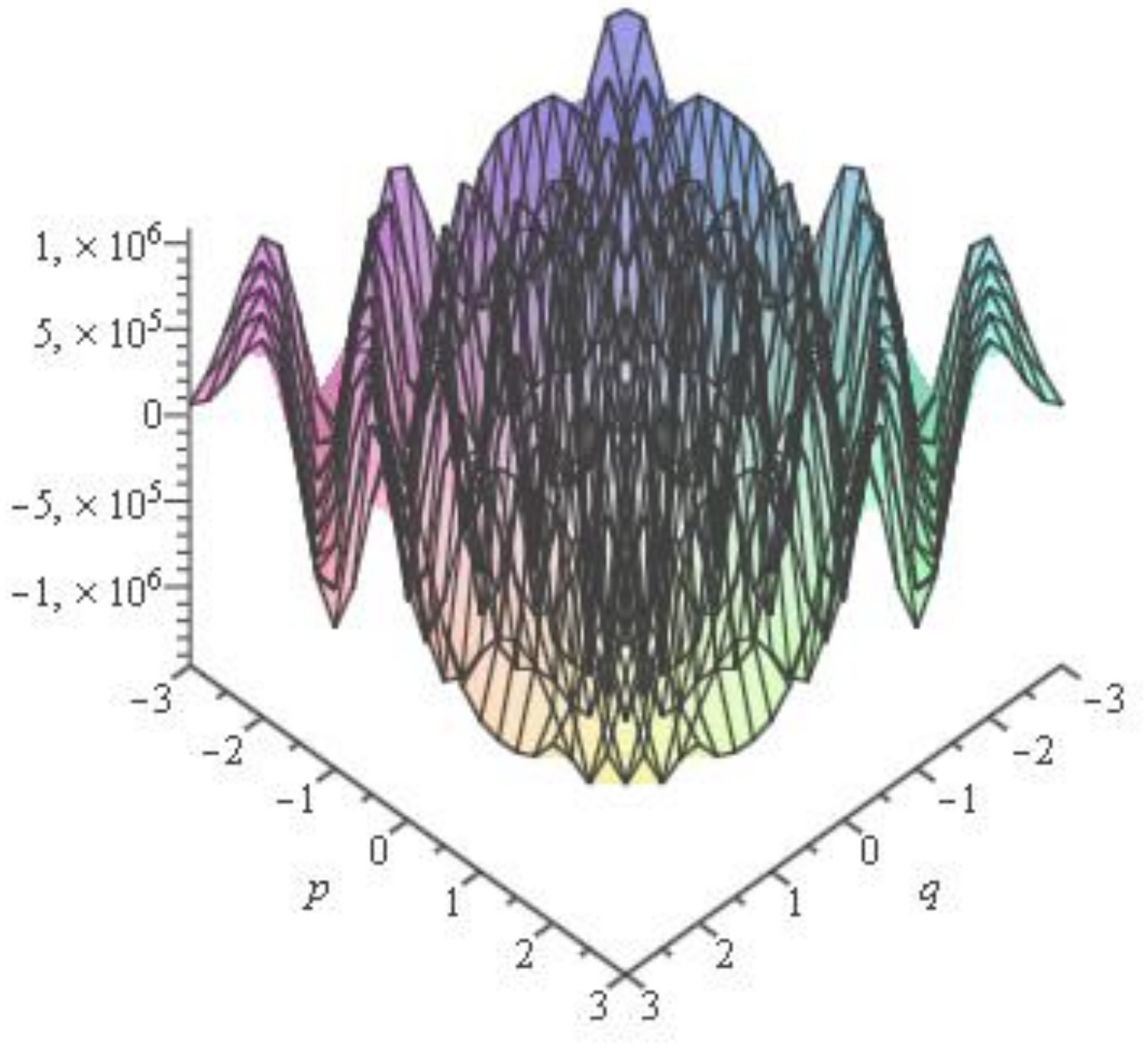}
\caption{Wigner function, $n=10$, $\lambda=0.9$} \label{figura 8}
\end{minipage}
\end{figure}

A measure of non-classicality of quantum states is defined on the volume of the negative part of Wigner function, which may be interpreted as a signature of quantum interference. In this sense, the non-classicality (negativity) indicator is given by~\cite{kenfack1}
\begin{eqnarray}\label{d12}
\eta(\psi)&=&\int\int [|W_{\psi}(q,p)|-W_{\psi}(q,p)]dqdq\\ \nonumber
&=&\int\int |W_{\psi}(q,p)|dqdq -1.\nonumber
\end{eqnarray}
This indicator represents the doubled volume of the integrated part of the Wigner function. In sequence, we calculated numerically this indicator for damped oscillator. The results of this calculation are shown in Table 1 below. A surprising result is that the parameter $\eta(\psi)$ does not depend on $\lambda$.
\begin{center}
\begin{tabular}{|c|c|}
\hline $n$ & $\eta(\psi)$\\
\hline 0  & 0\\
\hline 1 & 0.4261226344263795\\
\hline 2 & 0.7289892587057898\\
\hline 3 & 0.9766730799293403\\
\hline 4 & 1.1913424288065964\\
\hline 5  & 1.3834384856692004\\
\hline 6 & 1.5588521972493026\\
\hline 7 & 1.7212933835545317\\
\hline 8 & 1.873265816082318\\
\hline 9 & 2.016572434609475\\
\hline
\end{tabular}

Table 1. The non-classicality indicator as a function of the order of the Wigner function, the parameter $n$.
\end{center}

\begin{figure}[!htb]
\includegraphics[width=\linewidth]{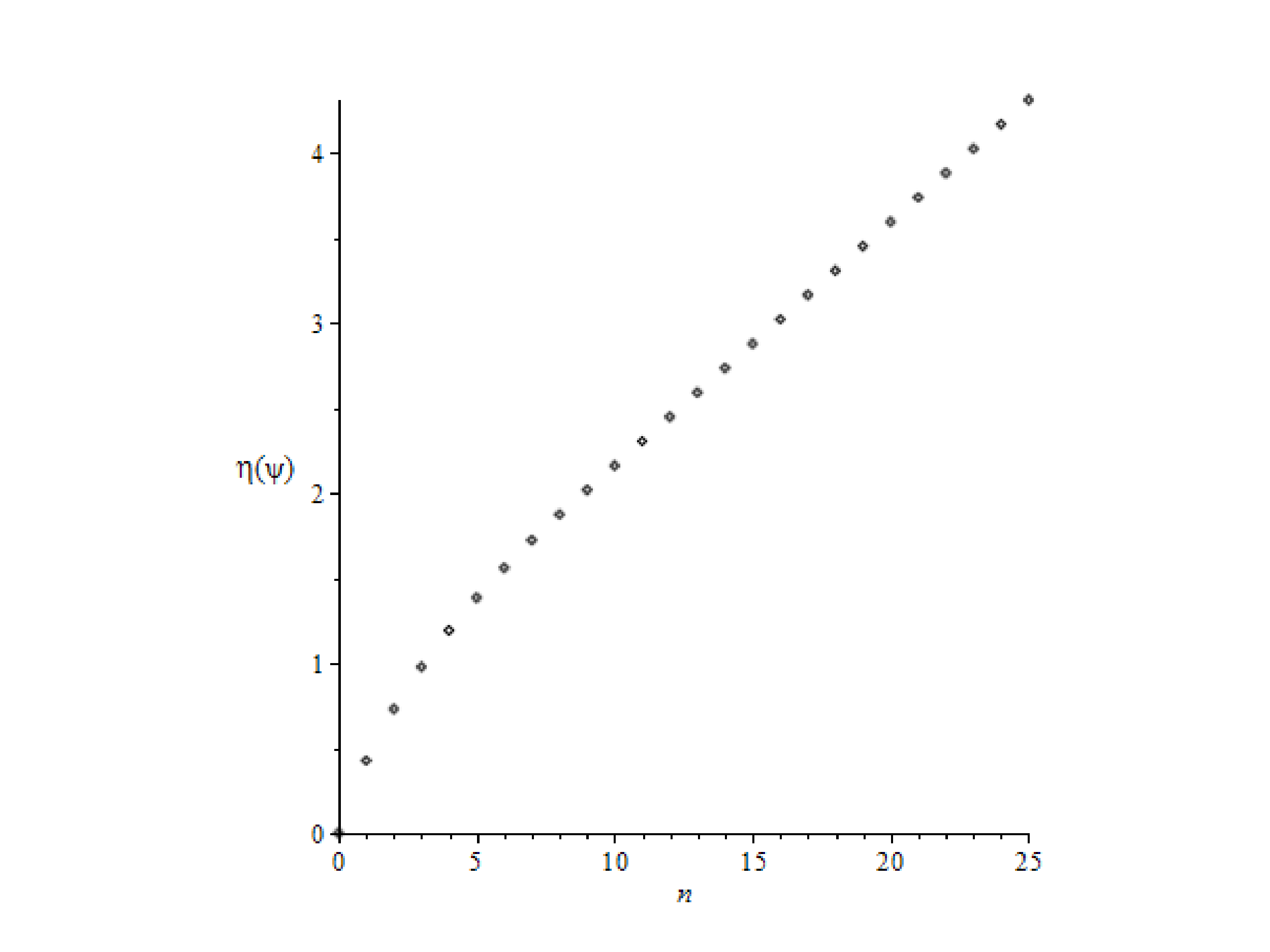}
\caption{The non-classicality indicator versus quantum number for damped oscillator $n\leq 50$} \label{figura 9}
\end{figure}

In Fig.~(\ref{figura 9}), the dependence of non-classicality indicator $\eta(\psi)$ and the order $n$ of Wigner function for damped oscillator is plotted. This allows us to conclude that the magnitude of the parameter $\lambda$, that represents the degree of damping, has no effect in the volume of the negative part of the  Wigner function.

\section{Concluding remarks}

In this work, the Wigner function for the  Helium atom is
calculated in the approximation of two-harmonic oscillators,
considering also dissipation.  Regarding the value of energy, this
approximation has provided satisfactory results with the
experiments~\cite{he1,he2,he3}. Here we have considered the
statistical nature of such quantum states, by analyzing the
non-classicality through the Wigner function. We have proceeded by
formulating the problem with the Schr\"{o}dinger equation in phase
space, such that the state, called a quasi-amplitude of
probability, is associated with the Wigner function by the Moyal
product.  In this context, we study a damped oscillator in phase
space. Using Wigner functions, a non-classicality indicator is
calculated as a function of the dissipation parameter. In this
case, the non-classicality behavior is independent of the
dissipation parameter.

\section*{Acknowledgements}
This work was partially supported by CNPq of Brazil.

\end{document}